\documentclass[pre,aps,epsfig,twocolumn,showpacs,amsmath,amssymb]{revtex4}
\usepackage{graphicx}
\usepackage{amssymb}
\usepackage{dcolumn}
\usepackage{color}
\begin{document}

\draft
\title{Condensation phenomena of conserved-mass aggregation model on weighted complex networks}
\author{Sungchul Kwon, Sooyeon Yoon and Yup Kim}
\affiliation{Department of Physics and Research Institute for Basic
Sciences, Kyung Hee University, Seoul 130-701, Korea}

\date{\today}

\begin{abstract}
We investigate the condensation phase transitions of
conserved-mass aggregation (CA) model on weighted scale-free
networks (WSFNs). In WSFNs, the weight $w_{ij}$ is assigned to the
link between the nodes $i$ and $j$. We consider the symmetric
weight given as $w_{ij}=(k_i k_j)^\alpha$. In CA model, the mass
$m_i$ on the randomly chosen node $i$ diffuses to a linked
neighbor of $i$,$j$, with the rate $T_{ji}$
 or an unit mass chips
off from the node $i$ to $j$ with the rate $\omega T_{ji}$. The
hopping probability $T_{ji}$ is given as $T_{ji}=
w_{ji}/\sum_{<l>} w_{li}$, where the sum runs over the linked
neighbors of the node $i$. On the WSFNs, we numerically show that
a certain critical $\alpha_c$ exists below which CA model
undergoes the same type of the condensation transitions as those
of CA model on regular lattices. However for $\alpha \geq
\alpha_c$, the condensation always occurs for any density $\rho$
and $\omega$. We analytically find $\alpha_c = (\gamma-3)/2$ on
the WSFN with the degree exponent $\gamma$. To obtain $\alpha_c$,
we analytically derive the scaling behavior of the stationary
distribution $P^{\infty}_k$ of finding a walker at nodes with
degree $k$, and the probability $D(k)$ of finding two walkers
simultaneously at the same node with degree $k$. We find
$P^{\infty}_k \sim k^{\alpha+1-\gamma}$ and $D(k) \sim
k^{2(\alpha+1)-\gamma}$ respectively. With $P^{\infty}_k$, we also
show analytically and numerically that the average mass $m(k)$ on
a node with degree $k$ scales as $k^{\alpha+1}$ without any jumps
at the maximal degree of the network for any $\rho$ as in the SFNs
with $\alpha=0$.

\end{abstract}

\maketitle

\section{Introduction}
A wide variety of mass transport systems ranging from traffic flow
to polymer gels \cite{E2,LEC,M,Z,W,S,MRRGI,F} exhibit nonequilibrium
condensation phenomena. These include basic microscopic dynamics
ubiquitous in nature such as aggregation, fragmentation and
diffusion. The nonequilibrium steady states of these systems are
classified into two types of phases, so-called fluid phase and
condensed phase. A finite fraction of total particles condenses on a
single site in the condensed phase. In the fluid phase, the particle
number on each site fluctuates around the density of total particles
($\rho$) without the condensation. As the rates of these processes
vary, the condensation phase transitions between the two phases may
take place at a certain critical density $\rho_c$.

One of the simplest mass transport models exhibiting the
condensation transitions is a conserved-mass aggregation (CA) model
\cite{sca,exactsca,aca,ca-mass,caKwon}. CA model evolves via
diffusion, chipping and aggregation upon contact which arise in a
variety of phenomena such as polymer gels \cite{Z}, the formation of
colloidal suspensions \cite{W}, river networks \cite{S,MRRGI} and
clouds \cite{F}. In one-dimensional CA model, the mass $m_i$ of site
$i$ moves either to site $i-1$ or to site $i+1$ with unit rate, and
then $m_i \rightarrow 0$ and $m_{i\pm1} \rightarrow m_{i\pm1}+m_i$.
With rate $\omega$, unit mass chips off from site $i$ and moves to
one of the nearest neighboring sites; $m_i \rightarrow m_i -1$ and
$m_{i\pm1}\rightarrow m_{i\pm1}+1$. The generalization to higher
dimensions is straightforward. As total masses are conserved, the
conserved density $\rho$ and the rate $\omega$ determine the phase
of CA model. The $\omega=\infty$ case corresponds to the well-known
zero range process (ZRP) with a constant hopping rate
\cite{zrp,NSL,zrp-w}.

The existence of the condensation transitions in CA model depends on
the symmetry of movement, the constraints of diffusion rate and the
underlying network structure \cite{exactsca,aca,ca-mass,caKwon}.
In the symmetric CA (SCA) model \cite{sca,exactsca} in which
diffusion and chipping direction are unbiased, the condensation
transitions take place at a certain $\rho_c$. The steady state
properties of SCA model is exactly described by the mean field
theory \cite{exactsca}. The single site mass distribution $P(m)$ was
shown to undergo phase transitions on regular lattices \cite{sca}.
For a fixed $\omega$, as $\rho$ is varied across the critical
density $\rho_c (\omega)$, the behavior of $P(m)$ for large $m$ was
found to be \cite{sca}

\begin{equation}\label{P}
 P(m)\sim
 \begin{cases}
          e^{-m/m^{*}}& \rho<\rho_c (\omega), \\
          m^{-\tau}& \rho=\rho_c (\omega), \\
          m^{-\tau}+ \text{\rm infinite\,\,\, aggregate}& \rho>\rho_c
          (\omega).
        \end{cases}
\end{equation}
Mean field theory predicts $\rho_c (\omega) = \sqrt{\omega+1}-1$ and
$\tau=5/2$ \cite{sca,exactsca}.

Recently, CA model on unweighted scale-free networks (SFNs) with
degree distribution $P(k)\sim k^{-\gamma}$ was studied to
investigate the effect of underlying network structure on the
condensation transitions \cite{caKwon}. We call networks with equal
weight on all links unweighted networks. On unweighted SFNs, the
same type of the condensation transitions as those of SCA in regular
lattice take place for $\gamma>3$. However for $\gamma \leq 3$, the
condensation always occurs for any density $\rho$($>0$). It was
shown that the existence of the transitions is directly related to
the diffusive capture process on unweighted SFNs
\cite{caKwon,dcpLee}.

On the other hand, most real-world networks exhibit not only a
heterogeneous distribution of degree, but also heterogeneous
distribution of weights \cite{wnet,scn,wan}. Weights assigned on
links characterize the interaction strengths between nodes. There
have been various attempts to understand the underlying mechanism
and scale-free behaviors of empirically observed weighted networks
\cite{wsfn}. Also there have been attempts to understand the effect
of heterogeneous weights on various dynamics such as
synchronization, dynamics of random walks, transport and
percolation, and condensation of zero-range process
\cite{scyn,rw-w,trans}. These studies showed that dynamical
properties are modified and exhibit non-trivial dependence on the
strength of weight. In this paper, as the generalization of our
study on CA model on complex networks, we investigate the effect of
both heterogeneous degree and weight on the condensation phenomena
of CA model on weighted networks.

The weight $w_{ji}$ represents the weight to a link from the node
$i$ to $j$. In general, the strength $s_i$ of the node $i$ scales
with the degree $k_i$ as $s_i \sim k_i^{\alpha}$. The exponent
$\alpha$ varies with network structures \cite{wsfn,wan}. Thus it
is natural to take the weight $w_{ji}$ as $w_{ji} \sim s_i s_j
\sim (k_i k_j)^{\alpha}$.

In this paper, we study the condensation transitions of CA on the
WSFNs with degree distribution $P(k) \sim k^{-\gamma}$ and the
symmetric weight $w_{ji} = (k_i k_j )^{\alpha}$. As in one
dimension, the diffusion of the whole masses and the fragmentation
of unit mass occur with the unit rate and the rate $\omega$,
respectively. In addition, masses move from a node $i$ to $j$ with
hopping rate proportional to $w_{ji}/\sum_{j} w_{ji}$. We found
that a certain critical $\alpha_c$ exists below which the
condensation transitions take place. However for $\alpha \geq
\alpha_c$, the condensation always occurs for any density $\rho
> 0$. To find $\alpha_c$ as a function of the degree exponent
$\gamma$, one needs the steady state distribution $P^{\infty}_k$ of
finding a walker at nodes with degree $k$ on the WSFNs.
$P^{\infty}_k$ gives the capture probability $D(k)$ with which two
walkers meet at a node with degree $k$. We analytically derived
$P_k$ and $D(k)$, and finally obtained $\alpha_c = (\gamma -3)/2$.

This paper is organized as follows. In Sec. II, we discuss the
condensation transitions of CA model on the WSFNs. To verify the
existence $\alpha_c$, we investigate the steady state property of a
single walker and the diffusive capture process on the WSFNs in Sec.
III and IV. We discuss the behavior of an average mass $m(k)$ of a
node with degree $k$ in Sec. V and finally summarize our results in
Sec. VI.

\section{ CA model on WSFNs with symmetric weights}
We consider CA model on WSFNs with the weight $w_{ij}$ from node
$j$ to $i$ defined as $w_{ij}=(k_i k_j )^{\alpha}$. For the
construction of WSFN, we first construct an unweighted static SFN
with $N$ nodes and $K$ links \cite{sfn}. The degree $k_i$ of a
node $i$ is defined as the number of its links connected to other
nodes. The average degree of a node $<k>$ is given as $<k>=2K/N$.
The degree distribution $P(k)$ of SFNs is a power-law distribution
$P(k)\sim k^{-\gamma}$. In the static model \cite{sfn}, it is
desired to use large $<k>$ to construct fully connected networks.
In simulations, we use $<k>=4$. After then, we assign a weight
$w_{ij}=(k_i k_j )^\alpha$ to the link between node $i$ and $j$.
Thus the hopping probability of masses from node $i$ to an $i'$s
linked neighbor $j$ is $T_{ji}=k^{\alpha}_i k^{\alpha}_j
/\sum_{<m>}k^{\alpha}_m k^{\alpha}_i= k^{\alpha}_j
/\sum_{<m>}k^{\alpha}_m$. $\sum_{<m>}$ denotes the sum over the
linked neighbors of node $i$.

Each node has an integer number of particles, and the mass on a node
is defined as the number of particles on the node. Initially $M$
particles are randomly distributed on $N$ nodes with given conserved
density $\rho = M/N$. Next a node $i$ is chosen at random and one of
the following events occurs:

(i) Diffusion : With the unit rate, the whole mass $m_i$ of node $i
$ moves to one of the linked neighbors $j$ with probability
$T_{ji}$. Then the aggregation takes place; $m_i \to 0$ and
$m_j\rightarrow m_j+m_i$.

(ii) Chipping : With the rate $\omega$, unit mass moves to a linked
neighbor $j$ with the probability $T_{ji}$, and then the aggregation
takes place, i.e. $m_i \to m_i -1$, $m_j \to m_j +1$.

The $\omega=\infty$ case corresponds to ZRP with constant chipping
rate on WSFNs \cite{zrp-w}.

We perform Monte Carlo simulations with random initial mass
distribution on the WSFNs with $\gamma=2.7$ and $4.0$. We set
$\omega=1$ and the network size $N=10^5$ with $<k>=4$. We measure
the single node mass distribution $P(m)$ in the steady states.

In Fig. 1, we plot $P(m)$ for $\gamma=4.0$ with two different
$\alpha$, $\alpha=0.05$ and $1.0$. $P(m)$ exhibits quite different
behavior according to the value of $\alpha$. For $\alpha=0.05$ (Fig.
1(a)), $P(m)$ decays exponentially without aggregates for
sufficiently low density $\rho=0.15$. On the other hand, for
sufficiently high density, $\rho=3.0$, an aggregate forms with the
power-law decaying background mass distribution. It means that the
condensation transition takes place at a certain critical density
$\rho_c (>0)$. Hence $P(m)$ follows Eq. (\ref{P}). Since in
unweighted SFNs, i.e $\alpha=0$, the condensation phase transitions
take place for $\gamma >3$ \cite{caKwon}, one may expect the
condensation transitions for very small $\alpha$. Based on the
following steps, we estimate $\rho_c$ and the exponent $\tau$.

In the condensed phase, the total density $\rho$ is written as
$\rho= \rho_c + \rho_\infty$, where $\rho_\infty$ is the density of
an aggregate. Since $\rho$ is given as $\rho=\int_1^\infty m P(m) dm
$, one can estimate $\rho_c$ from $\rho_c = \int_1^{m_o} mP(m) dm$,
where the upper bound $m_o$ is the cut-off mass at which the
background distribution terminates. Using this method, we estimate
$\rho_c =0.218$. We estimate the exponent $\tau$ from the scaling
plot $m^\tau P(m)$ using $P(m)$ of $\rho=3.0$ (Inset of Fig.1(a)).
Since the background distribution does not change for $\rho \geq
\rho_c$, we use $P(m)$ of $m \leq m_o$ for the scaling plot. We
estimate $\tau=2.38(5)$.

On the other hand, for $\alpha=1.0$ (Fig.1(b)), $P(m)$ shows the
complete different behavior. The condensation takes place with an
exponentially decaying background distribution for both sufficiently
low and high density, $\rho=0.1$ and $3.0$. Therefore we conclude
that the condensation always occurs for any nonzero density so a
system is always in the condensed phase without any transitions for
$\alpha=1.0$. The two different behaviors of $P(m)$ for
$\alpha=0.05$ and $1.0$ indicate that a crossover $\alpha$
($\alpha_c$) should exist in the range $0.05 <\alpha<1.0$ for
$\gamma=4.0$. A system undergoes the condensation transition for
$\alpha<\alpha_c$, while the condensation always occurs without the
transition for $\alpha \geq \alpha_c$.

\begin{figure}
\includegraphics[width=8cm]{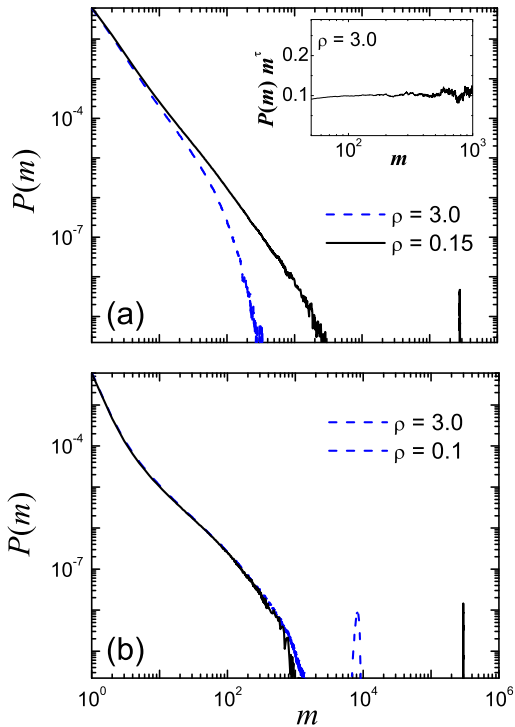}\label{fig1}
\caption{(Color online) The plot of $P(m)$ for $\gamma=4.0$ with
$\alpha=0.05$ (a) and $\alpha=1.0$ (b). The inset of (a) shows the
scaling plot $m^{\tau}P(m)$ with $\tau=2.38$ when $\rho=3.0$.}
\end{figure}

\begin{figure}
\includegraphics[width=8cm]{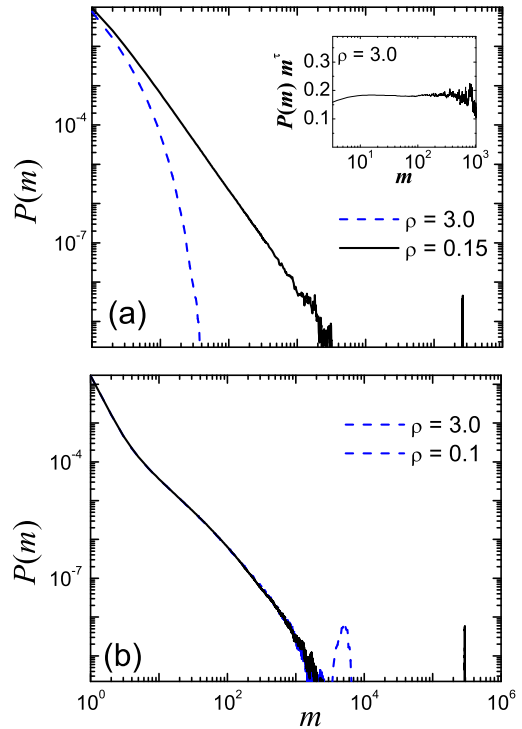}\label{fig2}
\caption{(Color online) The plot of $P(m)$ for $\gamma=2.7$ with
$\alpha=-1.0$ (a) and $\alpha=-0.05$ (b). The inset of (a) shows the
scaling plot $m^{\tau}P(m)$ with $\tau=2.46$ when $\rho=3.0$. }
\end{figure}

Similarly, for $\gamma=2.7$, $P(m)$ exhibits the same different
behavior according to the value of $\alpha$. The difference from the
$\gamma=4.0$ case is that the condensation transitions are observed
for a negative $\alpha$. We observe the condensation transitions for
$\alpha = -1.0$ (Fig. 2(a)). With the same method used in the
$\gamma=4.0$ case, we estimate $\rho_c =0.4$ and $\tau=2.46(5)$
respectively. However, for $\alpha=-0.05$, the condensation is
observed even for very low density $\rho=0.1$, which means that a
system is always in the condensed phase for $\alpha=-0.05$ (Fig.
2(b)). Therefore, the crossover $\alpha_c$ also exists for
$\gamma=2.7$, but its value is negative unlike the $\gamma=4.0$
case. Together with the results of $\gamma=4.0$, we conclude that
the crossover $\alpha_c$ exists for any $\gamma (>2)$ and $\alpha_c$
varies with $\gamma$. In what follows, we discuss the existence of
$\alpha_c$ and next the condensation nature for $\alpha<\alpha_c$.

First, the condensation phenomena of CA model on WSFNs is similar to
that on unweighted SFNs. On unweighted SFNs, the condensation
transitions exist for $\gamma>3$, while the condensation always
occurs for $\gamma \leq 3$ \cite{caKwon}. Hence the crossover
$\gamma$ is $\gamma_c =3$. Intriguingly, it was shown that the
existence of transitions is determined by the survival probability
of a diffusing prey chased by a diffusing predator, so-called the
lamb-lion problem \cite{lamb}. The reason is as what follows.

In the limit $\rho \to 0$, let us assume only an infinite aggregate
exists. With the rate $\omega$, the unit mass is chipped off and
moves around with the unit rate. If the chipped mass meets again
with the infinite aggregate within a finite time interval, then the
infinite aggregate is stable against the chipping process. On the
other hand, if the chipped mass and the infinite aggregate does not
meet again within a finite time interval, then the infinite
aggregate would disappear by repeated chipping processes. Therefore,
the stability of the infinite aggregate is physically related to the
capture process in which a diffusing lion (infinite aggregate)
chases a diffusing lamb (chipped mass). For the unweighted SFNs with
$\gamma \leq 3$, it was shown that the survival probability $S(t)$
of a lamb decays exponentially with finite life time $\langle T
\rangle_\infty$ \cite{caKwon,dcpLee}. However, for $\gamma
>3$, $S(t)$ is finite in the thermodynamic limit.
The behavior of $S(t)$ implies that the condensation transition
exist for $\gamma>3$ due to the stable fluid phase in the limit
$\rho \to 0$, but only the condensation exist for $\gamma \leq 3$.
As a result, the asymptotic behavior of the survival probability of
a lamb in the lamb-lion capture process determines the existence of
the condensation transitions on unweighted SFNs.

Similarly, on the WSFNs, the existence of the condensation
transitions is also expected to depend on the survival probability
$S(t)$ of a lamb. To see this, let us consider two limits, $\alpha
\to +\infty$ and $-\infty$ for a given $\gamma$. In the limit
$\alpha \to +\infty$, a walker always moves to a node with the
larger degree. Once a walker reach the hub node with the maximal
degree, the walker is trapped at the hub node forever. As a
result, a lion always captures a lamb at the hub node within a
finite time interval. Hence, $S(t)$ should decay exponentially
with a finite life time. On the other hand, in the limit $\alpha
\to - \infty$, a walker is forced to reach nodes with the minimal
degree. Due to the inhomogeneous structure, the nodes with the
minial degree are connected by nodes with larger degree. Hence, a
walker cannot escape from one of the nodes with the minimum degree
in this limit. It means that a lion cannot always capture a lamb
at some other node, so that $S(t)$ is finite.

From the behavior of $S(t)$ in the two opposite limits, there should
be a crossover $\alpha_c$. $S(t)$ is finite for $\alpha < \alpha_c$
and decays to zero for $\alpha \geq \alpha_c$. For the condensation
phenomena, one expects no condensation transitions ($\rho_c = 0$)
for $\alpha \geq \alpha_c$ due to finite life time of a lamb.
Instead, the condensation always occurs. On the other hand, the
condensation transitions occur for $\alpha < \alpha_c$. We
analytically find $\alpha_c = (\gamma-3)/2$ for a given $\gamma$ in
Sec. IV. From $\alpha_c = (\gamma-3)/2$, one reads $\alpha_c =
-0.15$ for $\gamma=2.7$ and $\alpha_c = 0.5$ for $\gamma=4$
respectively. Our simulation results for $\gamma=4$ and $2.7$
confirm the existence of $\alpha_c$ and also the sign of $\alpha_c$
for each $\gamma$.

Next, we discuss the critical behavior of CA model on WSFNs. The
CA model on any dimensional regular lattice and unweighted SFNs
with $\gamma>3$ is well described by mean-field theory
\cite{exactsca,caKwon}. On WSFNs, interestingly, the transitions
take place even for $\gamma<3$, which means that the transition
nature is not affected by the inhomogeneity of network structure.
Since $\alpha_c$ diverges for $\gamma \to \infty$, the critical
behavior of CA model on SFNs with $\alpha < \alpha_c$ should be
the same as that on random networks where $\alpha$, i.e. weight,
does not have no special meaning due to the uniform degree
distribution. As a result, one expects the mean-field critical
behavior of SCA model in regular lattice. Our numerical estimates
of $\tau$, $\tau=2.38(5)$ for $\gamma=2.7$ and $\tau=2.46(5)$ for
$4.0$, well agree with the mean-field value $\tau=5/2$. Therefore,
we conclude that the critical behavior of CA model for $\alpha <
\alpha_c$ on the WSFNs belongs to the universality class of SCA
model in regular lattice.

In summary, for a fixed $\gamma$, there is a crossover weight
exponent $\alpha_c$. CA model undergoes the same type of
condensation transitions as those of SCA model in regular lattice
for $\alpha<\alpha_c$, while the condensation always takes place for
nonzero density for $\alpha \geq \alpha_c$. To find $\alpha_c$ as a
function of the degree exponent $\gamma$, one needs the steady state
distribution $P^{\infty}_k$ of finding a walker at nodes with degree
$k$ on the WSFNs. In the next section, we derive $P^{\infty}_k$ on
the WSFNs. In Sec. IV, we study lamb-lion capture process on the
WSFNs and finally find $\alpha_c$ using $P^{\infty}_k$.

\section{Walks on WSFNs with symmetric weights}\label{brw}
We consider a single walker on weighted networks with the weight
$w_{ij}$. The connectivity of the network is represented by the
adjacency matrix $\mathbf{A}$ whose element $A_{ij}=1$ if there is
a link from a node $j$ to $i$. Otherwise, $A_{ij}=0$. We set
$A_{ii}=0$ conventionally. The degree $k_i$ of a node $i$ is given
as $k_i = \sum_j A_{ji}$. Since we consider weighted networks with
weight $w_{ij}$, we define the weighted adjacency matrix
$\widetilde{ \mathbf{A}}$ as $\widetilde{A}_{ij}=w_{ij}A_{ij}$.

The motion of a walker on the weighted networks defined by the
matrix $\widetilde{ \mathbf{A}}$ is a stochastic process in the
discrete time. We derive the stationary distribution $P^{\infty}_i$
of a walker being at node $i$ following the method of Ref.
\cite{noh}. To set up the equation, we define the transition
probability as follows. A walker at node $i$ at time $t$ selects one
of its $k_i$ linked nodes with hopping probability $T_{ji}$. Then,
at time $t+1$, the walker moves to the selected node. The hopping
probability $T_{ji}$ from node $i$ to $j$ is then given as
$T_{ji}=\widetilde{A}_{ji} / \widetilde{K}_i$, where
$\widetilde{K}_i = \sum_j \widetilde{A}_{ji}$ is the strength of
node $i$. As an initial condition, assume that the walker starts at
the node $q$ at time $t=0$. Then the recurrence relation of the
transition probability $P_{iq}$ of finding the walker at node $i$ at
time $t$ is
\begin{equation}\label{master1}
P_{iq}(t+1) = \sum_l T_{il} P_{lq}(t)\;\;.
\end{equation}
Then the transition probability $P_{iq}(t)$ is written by iterating
as

\begin{equation} P_{iq}(t)=\sum_{l_1, \cdots,
l_{t-1}} T_{i l_{t-1}} \cdots T_{l_2 l_1} T_{l_1
q}\;\;.\label{master2}
\end{equation}
For a symmetric $\widetilde{ \mathbf{A}}$ with $\widetilde{A}_{ij}
= \widetilde{A}_{ji}$, one finds $\widetilde{K}_q P_{iq}(t) =
\widetilde{K}_i P_{qi}(t)$ by comparing $P_{qi}$ and $P_{iq}$. In
the stationary state, the probability $P^{\infty}_i$ of finding a
walker at node $i$ should be independent of initial starting
nodes, which gives $\widetilde{K}_i P^{\infty}_q = \widetilde{K}_q
P^{\infty}_i$. Summing up over $q$, one finds
\begin{equation}
\label{Pi} P^{\infty}_i = \widetilde{K}_i/\mathcal{N} \;\;,
\end{equation}
where $\mathcal{N} = \sum_{q=1}^N \widetilde{K}_q = \sum_{q=1}^N
\sum_{m=1}^N \widetilde{A}_{mq}$. In weighted networks with
symmetric weights, $P^{\infty}_i$ is proportional to the strength of
node $i$, i.e. the sum of the weights of the nearest neighboring
nodes. The same result was found in the recent study on the dynamics
of random walks on growing weighted networks \cite{rw-w}.

In this paper, we consider the symmetric weight $w_{ij}$,
\begin{equation} \label{w}
w_{ij}= (k_i k_j)^{\alpha}\;\;.
\end{equation}
For the weight (\ref{w}), $P^{\infty}_i$ is not given as a simple
form. Hence it is better to handle the distribution $P^{\infty}_k$
of finding a walker at nodes with degree $k$. Using Eq.~(\ref{Pi}),
one can see that
\begin{equation}\label{Pk1}
P_k^\infty = \sum_{i=1}^{N} P^{\infty}_i \delta_{k_ik} =
\frac{1}{\mathcal{N}}\sum_{i=1}^N \sum_{j=1}^N
A_{ji}(k_ik_j)^\alpha\delta_{k_i,k}
\end{equation}
To express the sum in Eq.~(\ref{Pk1}) in terms of degree $k$, we
arrange the sum as follows. Only terms with $k_i = k$ contributes
nontrivially to the sum $\sum_{i}$ and thus $NP(k)$ nodes with the
degree $k$ in a network have the nontrivial contributions to the
sum. The node with the degree $k$ has $k$ linked neighbors whose
degree ranges from $1$ to the maximal degree of the network
$k_{max}$. Hence, the number of nontrivial terms in the sum
$\sum_{i=1}^N \sum_{j=1}^N$ is $NP(k)k$, which can be arranged in
the order of increasing degree. Then, the double sum of
Eq.~(\ref{Pk1}) is written as
$NP(k)k^{\alpha+1}(g(1)1+g(2)2^\alpha + \cdots +
g(k_{max})k^{\alpha}_{max})$, where $g(k')$ is the degree
distribution of the node involved in such $NP(k)k$ terms. For
large $N$, we approximate $g(k')$ to $P(k')$. Then $P^{\infty}_k$
is approximately given as
\begin{eqnarray}\label{Pk2}
P^{\infty}_k
&=&\frac{N}{\mathcal{N}}P(k)k^{\alpha+1}\int_{k_0}^{k_{max}}P(k')k'^{\alpha}
dk' \cr &=& P(k)k^{\alpha +1}/\int_{k_0}^{k_{max}}P(k')k'^{\alpha+1}
\;dk'\;\;.
\end{eqnarray}
On SFNs with degree distribution $P(k)\sim k^{-\gamma}$, the
integral in the second line is finite for $\alpha<\gamma-2$. Hence
we finally obtain $P^{\infty}_k$ on WSFNs as
\begin{equation}\label{Pk}
P_k^\infty \sim k^{\sigma}\;\;,\;\; \sigma = \alpha+1-\gamma\;\;.
\end{equation}
The exponent $\sigma$ varies with $\alpha$ and $\gamma$, and also
changes its sign. For $\alpha = \gamma -1$, i.e. $\sigma=0$,
$P_k^\infty$ is independent of degree $k$ so a walker does not feel
the inhomogeneity of the underlying network structure. While a
walker performs biased walks to nodes with the larger degree for
$\sigma>0$, the direction of the bias is reversed for $\sigma<0$.
Since the exponent $\alpha$ is a free parameter, one can controls
the direction of the bias for a given $\gamma$.

\begin{figure}
\includegraphics[width=8.0cm]{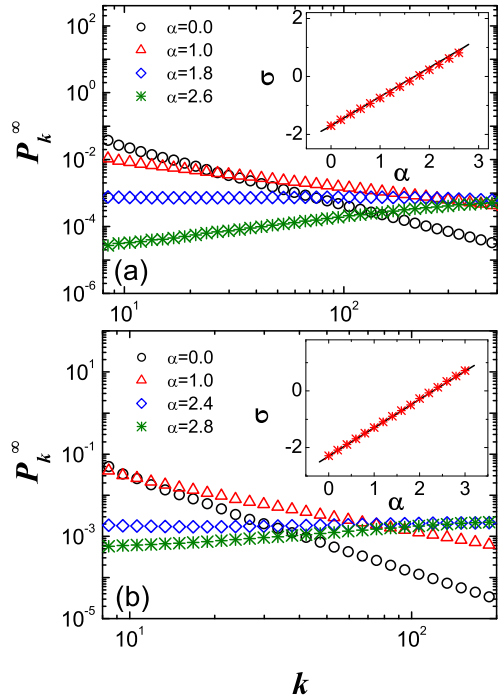}\label{fig3}
\caption{(Color online) The plot of $P^{\infty}_k$ and $\sigma$ for
$\gamma=2.7$ (a) and $\gamma=3.3$ (b). Insets show the relation
(\ref{Pk}) (solid line) and numerical estimates of $\sigma$
(symbols). }
\end{figure}

To check the scaling relation (\ref{Pk}), we perform Monte Carlo
simulations on the WSFNs with $N=10^5$ and the average degree
$<k>=4$. In the steady states, we measure $P^{\infty}_k$ for various
$\alpha$ up to $2.6$ for $\gamma=2.7$ and $3.0$ for $\gamma=3.3$.
Fig. 3 shows the plot of $P^{\infty}_k$ against $k$ for several
values of $\alpha$. As shown, $P^{\infty}_k$ scales in power-law
with $k$. The inset in each panel shows the plot of $\sigma$ against
$\alpha$. The simulation results agree well with the analytical
prediction (\ref{Pk}).

\section{Capture process on WSFNs with symmetric weights}\label{bdcp}

In this section, we consider the capture process or the lamb-lion
problem on WSFNs with the symmetric weights (\ref{w}). A lamb and a
lion initially locate separately on randomly selected two nodes.
Then the probability $D(k)$ of finding two walkers at the same node
with degree $k$ at the same time is proportional to $(P^{\infty}_k
)^2$. From Eq. (\ref{Pk}), one gets
\begin{equation}
D(k)=(P^{\infty}_k )^2 / NP(k) \sim k^{\nu}
\end{equation}
with
\begin{equation}\label{Dk}
\nu = 2(\alpha+1)-\gamma \;\;.
\end{equation}
Then the probability $D$ of finding two walkers on the same node
with any degree is given as
\begin{equation}
D = \int_{k_0}^{k_{max}} D(k) dk \sim \int_{k_0}^{k_{max}}
k^{2(\alpha+1)-\gamma} dk\;\;.
\end{equation}
Since the upper bound $k_{max}$ diverges with $N$, the integral
$\int_{k_0}^{k_{max}} k^{2(\alpha+1)-\gamma} dk$ diverges for
$\alpha \geq (\gamma-3)/2$. Hence there exists a crossover value
$\alpha_c$ given as
\begin{equation}\label{ac}
\alpha_c = (\gamma-3)/2 \;\;.
\end{equation}
For $\alpha < \alpha_c$, the lamb survives indefinitely with a
finite probability. However, for $\alpha \geq \alpha_c$, the lion
captures the lamb with the unit probability. To check the scaling
relation (\ref{Dk}), we measure $D(k)$ on the WSFNs with $N=10^5$
and $<k>=4$. For $10^5$ trials, we count the number $n(k)$ of
capture events on nodes with the degree $k$. We obtain $D(k)$ to
divide $n(k)$ by total trials($10^5$). Fig. 4 shows the plot of
$D(k)$ for several values of $\alpha$, which scales well with $k$
in power-law. As shown in the insets of Fig. 4, numerical
estimates for $\nu$ satisfy the relation (\ref{Dk}) very well.

\begin{figure}
\includegraphics[width=8cm]{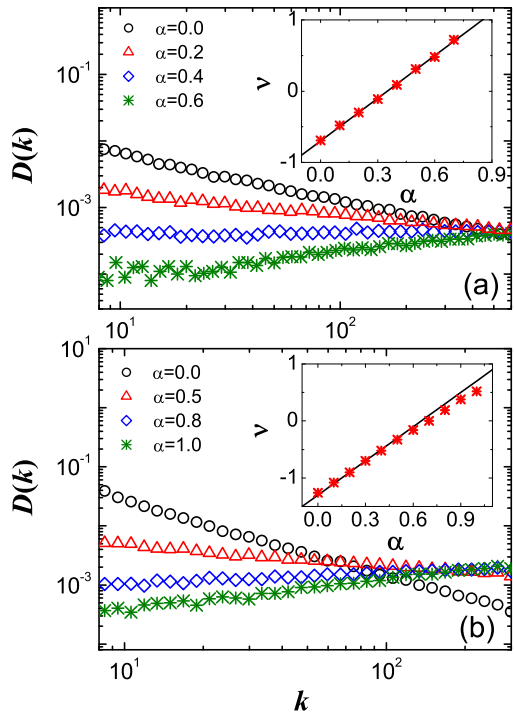}\label{fig4}
\caption{(Color online) The plot of $D(k)$ and $\nu$ for
$\gamma=2.7$ (a) and $\gamma=3.3$ (b). Insets show the relation of
(\ref{Dk}) (solid line) and numerical estimates of $\nu$ (symbols).
}
\end{figure}

\begin{figure}
\includegraphics[width=7.8cm]{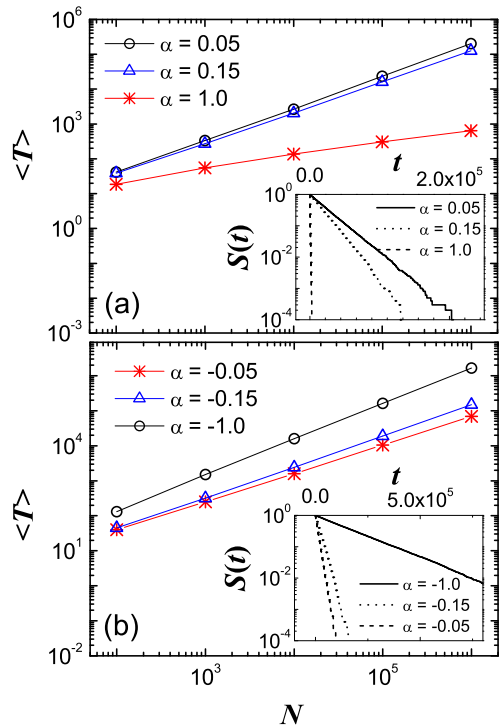}\label{fig5}
\caption{(Color online) The average lifetime $\langle T\rangle$ of a
lamb for $\gamma=3.3$ (a) and $\gamma=2.7$ (b). The solid line is a
guide to the eye. Insets show the semi-logarithmic plots of $S(t)$
for $N=10^5$. From top to bottom, each line corresponds to the
$S(t)$ of $\alpha<\alpha_c$, $\alpha_c$ and $\alpha > \alpha_c$
respectively.}
\end{figure}

To verify the existence of $\alpha_c$ by another method, we now
consider the survival probability $S(t)$ of a lamb. $S(t)$ always
satisfies $S(t)=S_\infty \;e^{-t/\tau}$ on random and scale-free
networks due to the small world nature \cite{caKwon,dcpLee}. As
$S(t)=S_\infty \;e^{-t/\tau}$ in SFNs with any $\gamma$, we are
interested in the average life time $\langle T\rangle$ of a lamb
rather than $S(t)$ itself. From $\langle T\rangle =
\int_{0}^{\infty} [-dS(t)/dt]\; dt$ and $S(t)=S_\infty \;
e^{-t/\tau}$, we have $\langle T\rangle \sim \tau$. Hence $\langle
T\rangle$ is infinite for $\alpha < \alpha_c$ and finite for
$\alpha \geq \alpha_c$ in the limit $N \to \infty$. However, for
the finite-sized networks, a lamb is eventually captured within
$N$ time steps for any $\alpha$. For $\alpha<\alpha_c$, the
maximum life time should be the order of $N$ to guarantee the
finite survival probability in the limit $N \to \infty$. Hence
$\langle T\rangle$ is expected to scale as $\langle T\rangle \sim
N$ for $\alpha<\alpha_c$. We measure $\langle T\rangle$ on WSFNs
of $\gamma=2.7$ and $3.3$ with network size $N$ up to $10^6$. From
Eq. (\ref{ac}), one reads $\alpha_c =-0.15$ for $\gamma=2.7$ and
$\alpha_c = 0.15$ for $\gamma=3.3$.

In Fig. 5, we plot $\langle T\rangle$ against $N$. As shown in
each inset, $S(t)$ exponentially decays for any $\alpha$. For
$\gamma=3.3$(Fig. 5(a)), $\langle T\rangle$ increases with $N$ as
$N^\phi$ with $\phi = 0.94(1)$ for $\alpha = 0.05$ ($<\alpha_c$)
and $\phi=0.90(1)$ for $\alpha=\alpha_c (=0.15)$. We estimate
$\phi$ by measuring successive slopes from the log-log data in
Fig. 5(a). For $\alpha=1.0$ ($> \alpha_c$), $\langle T\rangle$
tends to saturate to the asymptotic value $\langle
T\rangle_\infty$ with decreasing successive slopes. The exponent
$\phi$ of $\alpha=0.05$ is close to the expected value $\phi=1$.
For $\alpha=\alpha_c(=0.15)$, $\langle T\rangle$ seems to diverge
with $\phi=0.9$. However, since $\langle T\rangle$ of $\alpha=1.0$
already tends to saturate, it is expected that $\langle T\rangle$
of $\alpha \geq \alpha_c$ would saturate to a finite value in the
network with $N > N_c(\alpha)$, where $N_c(\alpha)$ is the
characteristic size for given $\alpha$. For example, for
$\alpha=1.0$, $\langle T\rangle$ does not get into the saturation
region even after $N=10^6$, which implies $N_c(1.0) >10^6$ for
$\alpha=1.0$. Since $N_c$ should increase as $\alpha \to
\alpha_c$, it is empirically impossible to see the saturation of
$\langle T\rangle$ via simulations. Therefore, the initial slope
$\phi$ at $\alpha_c$ may have no special meaning as that of
$\alpha > \alpha_c$. The same behavior for $\langle T\rangle$
 was observed for $\alpha=0$ case
\cite{caKwon}, where $\langle T\rangle$ initially algebraically
increases with continuously varying $\phi$($<1$) as $\gamma \to 3$
from below.

For $\gamma=2.7$ (Fig. 5(b)), we estimate $\phi=1.00(2)$ for
$\alpha=-1.0$ ($<\alpha_c$) as expected. However, for
$\alpha=-0.05$ ($>\alpha_c$), $\langle T\rangle$ algebraically
increases with $\phi=0.82(1)$. Since for $\alpha=0$ \cite{caKwon},
$N_c$ is already larger than $10^6$ for $\gamma=2.75$, it is
difficult to see the saturation of $\langle T\rangle$. For
$\alpha=\alpha_c$, we estimate $\phi=0.90(1)$. As in $\gamma=3.3$,
the initial slope for $\alpha \geq \alpha_c$ has no special
meaning. Based on our numerical results, we are convinced that
$\langle T \rangle$ approaches a finite value $\langle T
\rangle_\infty$ for $\alpha \geq \alpha_c$ and becomes infinite
for $\alpha<\alpha_c$ in the limit $N \to \infty$. Hence in the
limit $N \to \infty$, we have
\begin{equation}\label{S}
\lim_{N \rightarrow \infty} S(N,t) = \begin{cases}
S_o e^{-t/\tau_{\infty}} & (\alpha \geq \alpha_c) \\
S_{\infty} & (\alpha < \alpha_c)\;\;.
 \end{cases}
\end{equation}
with the characteristic time $\tau_\infty \sim \langle T
\rangle_\infty$.

\section{Average mass of a node with degree $k$} Another
interesting quantity in condensation phenomena on networks is the
average mass $m(k)$ at a node with degree $k$ in the steady state
\cite{caKwon,NSL,zrp-w}. In ZRP with chipping rate $u(m)\sim
m^{\delta}$, the complete condensation takes place for $\delta <
\delta_c$, where $\delta_c =1/(\gamma-1)$ for unweighted SFNs
\cite{NSL} and $(\alpha+1)/(\gamma-1)$ for WSFNs with the weight
(\ref{w}) \cite{zrp-w}. For $\delta < \delta_c$, $m(k)$ increases
as $k^{\alpha+1}$ for $k < k_c$, and as $k^{(\alpha+1)/\delta}$
for $k \geq k_c$ on the WSFNs. Especially for $\delta=0$, $m(k)$
increases $k^{\alpha+1}$ until $k < k_{max}$ and jumps to the
value $m_{hub} \approx \rho N$ at $k_{max}$. Hence the
condensation takes place at the node with $k_{max}$ degree in ZRP.

\begin{figure}
\includegraphics[width=9cm]{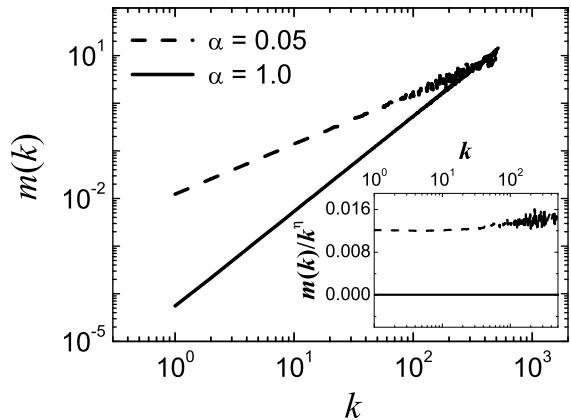}\label{fig6}
\caption{The plot of $m(k)$ for $\gamma=3.3$. The solid and the
dashed line correspond to $\alpha=1.0$ and $0.05$ respectively. The
inset shows the scaling plot $m(k)k^{-\eta}$ with $\eta=1.06$ for
$\alpha=0.05$ (dashed line) and $\eta=1.95$ for $\alpha=1.0$ (solid
line).}
\end{figure}

The recent study on CA model on unweighted SFNs showed that $m(k)$
linearly increases up to $k_{max}$ without the jump at $k_{max}$
unlike in ZRP with constant chipping rate \cite{caKwon}. The
linearity of $m(k)$ comes from the fact that all masses can
diffuse. The mass $m_{hub}$ formed on the node with the degree
$k_{max}$ can diffuse throughout network to make the steady sate
distribution $P^{\infty}_i$. By taking average over all nodes, the
$m_{hub}$ soaks into the average mass $m(k)$ unlike in ZRP where
all samples have $m_{hub}$ at $k_{max}$. The linearity of $m(k)$
on unweighted SFNs results from $P^{\infty}_k \sim kP(k)$
\cite{caKwon}. Therefore, from $P^{\infty}_k \sim k^{\alpha+1}
P(k)$ on the WSFNs, we expect $m(k) \sim k^{\alpha+1}$ up to
$k_{hub}$. To see this explicitly, we derive the relation $m_k
\sim k^{\alpha+1}$ as follows.

We consider the average total mass $M(k)$ of nodes with degree $k$
defined as
\begin{equation}\label{Mk}
M(k)= \sum_{m=0}^{\infty}mP_\infty (m,k)\;\;,
\end{equation}
where $P_\infty (m,k)$ is the probability of finding a walker with
mass $m$ at nodes with degree $k$ in the steady state. Since the
mass distribution $P(m)$ in the steady state is independent of $k$,
we write $P_\infty (m,k)=P(m)P_k^\infty$. From (\ref{Pk2}) and
(\ref{Mk}), one reads
\begin{equation}
M(k) \simeq k^{\alpha+1}P(k)\sum_{m=0}^{\infty} mP(m) \;\;,
\end{equation}
where we drop the normalization constant of $P^{\infty}_k$. Since
the number of nodes with degree $k$ is $NP(k)$, $m(k)$ is given as
\begin{equation}\label{mk}
m(k)=\frac{M(k)}{NP(k)} \sim k^{\alpha+1} \;\;.
\end{equation}

To confirm the scaling behavior of $m(k)$, we measure $m(k)$ in the
condensed phase on the WSFNs of $\gamma=3.3$ and $N=10^5$. In Fig.
6, we plot $m(k)$ against $k$ for $\alpha=0.05$ and $1.0$. With
$\omega=1$, we set $\rho=3.0$ which corresponds to the condensed
phase for both $\alpha$ values. Assuming $m(k)\sim k^\eta$, one
expects $\eta=1.05$ for $\alpha=0.05$ and $2.0$ for $\alpha=1.0$
respectively. From the scaling plot $m(k)/k^\eta$ (Inset of Fig. 6),
we estimate $\eta = 1.06(2)$ for $\alpha=0.05$ and $\eta=1.95(5)$
for $\alpha=1.0$ which agree well with the predictions.

\section{Summary}
In summary, we investigate the properties of conserved-mass
aggregation (CA) model on weighted scale-free networks (WSFNs). In
WSFNs, the weight $w_{ij}$ is assigned to the link between node
$i$ and $j$. We consider the symmetric weight given as
$w_{ij}=(k_i k_j)^\alpha$. In CA model, masses diffuse with unit
rate and unit mass chips off from mass with rate $\omega$. In
addition, the hopping probability $T_{ji}$ from node $i$ to $j$ is
given as $T_{ji}= w_{ji}/\sum_{<m>} w_{mi}$.

On the WSFNs, a walker finally reaches the hub node with $k_{max}$
degree for $\alpha \to \infty$, while it is trapped forever at
nodes with the minimal degree for $\alpha \to -\infty$. In the
lamb-lion capture process, it means that the lion captures the
lamb at the hub node within finite time interval for $\alpha \to
\infty$. On the other hand, a lamb survives indefinitely with
finite probability for $\alpha \to -\infty$, because the lion
cannot escape from a node with the minimal degree to capture a
lamb at some other node. In-between the two limits, one expects a
crossover $\alpha_c$ below which the life time $\langle T \rangle$
of a lamb is infinite. However, for $\alpha \geq \alpha_c$,
$\langle T \rangle$ is finite. The dependence of $\langle T
\rangle$ on $\alpha$ is similar to that on unweighted SFNs of
$\alpha=0$ where $\langle T \rangle$ is infinite for $\gamma >3$
and finite otherwise \cite{caKwon}.

To verify the existence of $\alpha_c$, we need the stationary
distribution $P^{\infty}_k$ of finding a walker at nodes with
degree $k$. From the  equation for the transition probability
$P_{ji} (t)$ to go from node $i$ to $j$ in $t$ time steps, we
analytically find $P^{\infty}_k \sim k^{\alpha+1-\gamma}$. Next,
we consider the so-called lamb-lion capture process. With
$P^{\infty}_k$, we find the probability $D(k)$ of finding two
walkers at the same node with degree $k$ at the same time to scale
$D(k) \sim k^{2(\alpha+1)-\gamma}$. Finally, integrating out
$D(k)$, we find the death probability $D$ of a lamb. A lamb
survives indefinitely with the finite survival probability for
$\alpha < \alpha_c$, while it is eventually captured by a lion for
$\alpha \geq \alpha_c$. We analytically find $\alpha_c =
(\gamma-3)/2$. Therefore, in the limit $N \to \infty$, the life
time $\langle T \rangle$ of a lamb is finite for $\alpha \geq
\alpha_c$, while it is infinite for $\alpha< \alpha_c$. We
numerically confirm the all analytical results.

The existence of the condensation transitions is known to depend on
$\langle T \rangle$ of a lamb \cite{caKwon}. For $\alpha \geq
\alpha_c$, $\langle T \rangle$ is finite so the condensation always
occurs for any nonzero density. On the other hand, for $\alpha
<\alpha_c$, the infinite $\langle T \rangle$ ensures the
condensation transitions at a certain critical density $\rho_c$. For
$\alpha \geq \alpha_c$, we numerically confirm that the condensation
always takes place at very low density. We also numerically confirm
that for $\alpha < \alpha_c$, CA model on the WSFNs undergoes the
same type of the condensation transitions as those of SCA model in
regular lattice.

Finally, we investigate the behavior of the average mass $m(k)$ of
a node with degree $k$. In ZRP with constant chopping rate on
networks \cite{NSL,zrp-w}, $m(k)$ increases as $k^{\alpha+1}$, and
jumps to the total mass of the system at $k_{max}$. However, in
the SCA model on unweighted SFNs, it was shown that $m(k)$
linearly increases with $k$ up to $k_{max}$ without any jumps
\cite{caKwon}. Furthermore the linearity of $m(k)$ is valid for
any $\rho >0$, which comes from the fact that the diffusion is
only the relevant physical factor to decide the distribution
$m(k)$. Similarly, on the WSFNs, we analytically find and
numerically confirm that $m(k)$ algebraically increases as
$k^{\alpha+1}$ for any $\rho>0$ without any jumps.

\begin{acknowledgments}
This work was supported by the Korea Science and Engineering
Foundation(KOSEF) grant funded by the Korea government(MOST) (No.
R01-2007-000-10910-0) and by the Korea Research Foundation Grant
funded by the Korean Government (MOEHRD, Basic Research Promotion
Fund) (KRF-2007-313-C00279).
\end{acknowledgments}


\begin{references}
\bibitem{E2} M. R. Evans, Europhys. Lett. 36, 13 (1996).
\bibitem{LEC} O. J. O'Loan, M. R. Evans, and M. E. Cates, Phys. Rev. E 58,
1404 (1998).
\bibitem{M} D. van der Meer et al., J. Stat. Mech. Theor. Exp. 04,
P04004 (2004); J. Torok, cond-mat/0407567.
\bibitem{Z} R. M. Ziff, J. Stat. Phys. 23, 241 (1980).
\bibitem{W} W. H. White, J. Colloid Interface Sci. 87, 204 (1982).
\bibitem{S} A. E. Scheidegger, Bull. I.A.S.H. 12, 15 (1967).
\bibitem{MRRGI} A. Maritan, A. Rinaldo, R. Rigon, A. Giacometti, and
I. R. Iturbe, Phys. Rev. E 53, 1510 (1996); M. Cieplak, A.
Giacometti, A. Maritan, A. Rinaldo, I. R. Iturbe, and J. R. Banavar,
J. Stat. Phys. 91, 1 (1998).
\bibitem{F} S. K. Friedlander, Smoke, Dust and Haze (Wiley
Interscience, New York, 1977).

\bibitem{sca} S. N. Majumdar, S. Krishnamurthy, and M. Barma,
Phys. Rev. Lett. 81, 3691 (1998); S. N. Majumdar, S. Krishnamurthy,
and M.Barma, J. Stat. Phys. 99, 1 (2000).
\bibitem{exactsca}R. Rajesh and S. N. Majumdar, Phys. Rev. E {\bf
63}, 036114 (2001).
\bibitem{aca} R. Rajesh and S. Krishnamurthy, Rhys. Rev. E 66, 046132(
2002).
\bibitem{ca-mass}R. Rajesh, D. Das, B. Chakraborty, and M. Barma,
Phys. Rev. E {\bf66}, 056104(2002).
\bibitem{caKwon} S. Kwon, S. Lee, and Yup Kim, Phys. Rev. E {\bf
73}, 056102 (2006).

\bibitem{zrp} M. R. Evans, Braz.J.Phys. 30, 42 (2000); M. R.
Evans and T. Hanney, J. Phys. A {\bf38}, R195 (2005).
\bibitem{NSL} J. D. Noh, G. M. Shim, and H. Lee, Phys. Rev. Lett. 94,
198701 (2005).
\bibitem{zrp-w} M. Tang, Z. Liu, and J. Zhou, Phys. Rev. E {\bf74},
036101 (2006).

\bibitem{wnet} S. L. Pimm, {\it Food Webs} (University of Chicago
Press. Chicago. 2002). 2nd Ed.; A. E. Krause et al, Nature (London)
{\bf 426}, 282 (2003).
\bibitem{scn} M. E. J. Newman, Phys. Rev. E {\bf64}, 016132 (2001).
\bibitem{wan} A. Barrat,M. Barth\'{e}lemy, R. Pastor-Santorras, and A. Vespignani,
 Proc. Natl. Acad. Sci. U.S.A. {\bf 101}, 3747
(2004).
\bibitem{wsfn} S. H. Yook, H. Jeong, A.-L. Barabasi, and Y. Tu,
Phys. Rev. Lett. {bf 85}, 5835 (2001); A. Barrat, M. Barthelemy, and
A. Vespignani, {\it ibid}, {\bf 92}, 228701 (2004); K.-I. Goh, B.
Kahng, and D. Kim, Phys. Rev. E {\bf 72}, 017103 (2006).

\bibitem{scyn} A. E. Motter, C. S. Zhou, and J. Kurths, Phys. Rev. E
{\bf71}, 016116 (2005).
\bibitem{rw-w} An-Cai Wu et al, Chin. Phys. Lett {\bf24}, 557
(2007).
\bibitem{trans} G. Li et al, Phys. Rev. E {\bf75}, 045103(R) (2007).

\bibitem{noh} J. D. Noh and H. Rieger, Phys. Rev. Lett. {\bf 92}, 118701
(2004).
\bibitem{dcpLee} S. Lee, S. -H. Yook, and Yup Kim, Phys. Rev. E {\bf
74}, 046118 (2006).
\bibitem{sfn} K.-I. Goh, B. Kahng, and D. Kim, Phys. Rev. Lett.
{\bf87}, 278701 (2001).
\bibitem{lamb} P. L. Krapivsky and S.Redner, J. Phys. A 29, 5347 (1996).
\end{references}
\end{document}